# Manuscript for

Distributional regression models for meta-analysis

**Authors:**

Yefeng Yang[1,2], Shinichi Nakagawa[2,3]

**Affiliation:**

[1]Biosystems Engineering, Zhejiang University, Hangzhou, 310058, China

[2]School of Biological, Earth and Environmental Sciences, University of New South Wales, Sydney, NSW 2052, Australia

[3]Department of Biological Sciences, University of Alberta, CW 405, Biological Sciences Building, Edmonton, AB T6G 2E9, Canada

**Replication package:**

The data and scripts needed to reproduce the analyses and figures are archived GitHub repository https://github.com/Yefeng0920/distMA.

**Abstract**

Meta-analyses are regarded as the highest level in the hierarchy of evidence, yet standard models traditionally concentrated on estimating the mean effect size, often under restrictive assumptions about the underlying distribution, such as homogeneous variance, symmetric shapes. We introduce a distributional regression framework for meta-analysis that generalizes these conventional models by allowing all parameters of the effect size distribution, such as location, scale, and shape, to be modelled as functions of explanatory variables. This unified framework accommodates a wide range of existing models, including random-effects, multilevel, multivariate, location-scale, and outlier-robust meta-analyses, as special cases. We provide an illustrative example, using 67,393 meta-analyses from the Cochrane Database of Systematic Reviews, employing location-scale models to investigate whether smaller studies tend to report larger effect sizes (i.e., small-study effects) and exhibit greater heterogeneity. We discuss implementation strategies using existing software, considerations for model selection and pre-registration, and the need for further methodological development. By moving beyond the mean effect size, distributional regression enables researchers to explore systematic variation in distributional structure, facilitating the joint test of new hypotheses corresponding to multiple distributional parameters.

# 1 Introduction

Meta-analysis is a cornerstone of quantitative evidence synthesis, providing a rigorous and reproducible statistical framework for integrating results from multiple studies to obtain more precise estimates of effect sizes and systematically assess sources of heterogeneity (Borenstein et al., 2009; Gurevitch et al., 2018). Well-designed and conducted meta-analyses are widely regarded as the highest level in the hierarchy of evidence, shaping clinical decision-making and informing policy recommendations (Ineichen et al., 2024). While traditionally dominant in clinical research, meta-analysis has become increasingly valuable in preclinical domains, encompassing *in vitro* experiments, animal models of human diseases, and physiological and behavioural studies (Yang et al., 2023). In some cases, preclinical meta-analyses also incorporate clinical studies on the same topic, bridging the gap between experimental and clinical research.

By synthesizing data from diverse experimental settings, preclinical meta-analysis plays a crucial role in identifying biological mechanisms, assessing the translational potential of findings, and improving reliability (Ineichen et al., 2024). Furthermore, it informs best practice guidelines, reduces research waste by highlighting inconsistencies and methodological shortcomings, and facilitates more efficient allocation of research resources. For example, in identifying biological mechanisms, a meta-analysis synthesizing data from 4,786 rodent studies demonstrated that parental exercise enhances offspring brain development by upregulating brain-derived neurotrophic factor (BDNF) and vascular endothelial growth factor (VEGF) (Yang et al., 2021). Beyond mechanistic insights, meta-analysis is also instrumental in assessing the translational relevance of preclinical findings. A meta-analysis comparing treatment effects across preclinical animal models and human clinical trials found that therapies tested in experimental conditions that closely mimic real-world patient populations, such as stroke interventions in aged animals with comorbidities, were more likely to show efficacy in clinical trials (Perel et al., 2007). However, despite the widespread application of preclinical meta-analysis, the reliability of meta-analytic evidence critically depends on the statistical assumptions underlying meta-analytic models. When these assumptions are violated—as is often in practice—statistical inferences can become biased or misleading, potentially undermining the credibility and utility of meta-analytic findings (Jackson & White, 2018).

Meta-analysis is, at its core, a statistical approach in modelling the distribution of outcomes (i.e., effect sizes such as odds ratios) across studies. Standard meta-analytic models—including fixed-effect and random-effects models, typically impose strong parametric distributional assumptions (e.g., constraints on location and scale parameters) (Borenstein et al., 2009). However, when these assumptions are violated, statistical inference may become unreliable: confidence intervals no longer maintain their nominal coverage, hypothesis tests yield misleading p-values, and overall conclusions drawn from meta-analyses may be systematically biased. To address these issues, various advanced models have been proposed, including generalized linear mixed models (GLMMs) for handling non-normal sampling distributions (Jackson et al., 2018; Y. Lee & Nelder, 2006; Stijnen et al., 2010), multilevel and multivariate meta-analysis models for accounting for statistical dependencies (Riley, 2009; Van den Noortgate et al., 2013), location-scale models for modelling heterogeneity as a function of study-level characteristics (Rodriguez et al., 2023; Viechtbauer & López-López, 2022), and robust meta-analysis methods that employ heavy-tailed distributions to accommodate outlier studies (Baker & Jackson, 2016). While these methods provide more flexible alternatives to conventional models, they remain separate approaches rather than components of a unified statistical framework. This fragmentation calls for the need for a more general and coherent framework that can accommodate the complex data structures and diverse distributional properties encountered in the real world.

We propose distributional regression as a unifying statistical framework for meta-analysis, extending conventional models by allowing not only the location parameter (i.e., mean) but also all other distributional parameters (e.g., scale and shape parameters) to vary across studies. Just as the random-effects model relaxes the assumption of homogeneous location (mean) parameters, distributional regression generalizes this concept by treating all distributional parameters as heterogeneous rather than fixed constants. Crucially, distributional regression models allow all distributional parameters to be flexibly modelled as functions of covariates or moderator variables that reflect biological, methodological, and sociological (or meta-scientific) characteristics of preclinical studies. Embedding meta-analysis within the distributional regression framework unlocks new opportunities for (pre-)clinical and other areas of meta-analytic research, enabling researchers to explore beyond mean effects and formulate new hypotheses from published meta-analyses. Additionally, distributional regression mitigates common model misspecifications, such as non-normal and correlated sampling errors, heteroscedasticity, zero inflation, and overdispersion, which may often undermine the validity of conventional meta-analytic models. To illustrate the potential of this framework, we apply distributional regression models to a dataset of over 60,000 meta-analyses from the Cochrane Database of Systematic Reviews (Bartoš et al., 2024). Our analysis revisits long-standing questions about publication bias and heterogeneity in meta-analysis and demonstrates how distributional regression can reveal new insights from existing data.

The structure of the paper is as follows. The next section outlines key distributional assumptions in meta-analytic models, highlighting their limitations. We then present the general form of the proposed distributional regression model for meta-analysis and how conventional and advanced meta-analytic models—including random-effects, multilevel, multivariate, location-scale, and robust models—can be expressed as special cases of this framework. Subsequently, we apply a distributional regression model to the Cochrane database and discuss the implications of our findings for meta-analytic practice. Finally, we conclude with a discussion of recommendations for incorporating distributional regression into routine meta-analysis.

## 2 Assumptions of standard meta-analytic models and extensions addressing their violations solutions

Meta-analysis in preclinical research frequently employs the random-effects (RE) model to synthesize effect sizes across studies, offering a simple framework for estimating an overall effect while accounting for heterogeneity (Vesterinen et al., 2014; Yang et al., 2023). To elucidate the assumptions underlying this model and their potential violations, we first present its mathematical formulation, followed by a discussion of its constraints and extensions that address these limitations (Table 1). This exposition aims to provide a clear foundation for understanding the distributional regression framework proposed in subsequent sections, particularly in the context of preclinical data, where small sample sizes and diverse experimental conditions often challenge standard assumptions.

Consider a meta-analysis comprising $k$ independent studies, each contributing a single effect size estimate $y_i$, such as a standardized mean difference or log-odds ratio, for study $i = 1, ..., k$. The RE model posits that the observed effect size $y_i$ is the sum of the true effect size parameter $\mu_i$ and Gaussian error:

$$y_i = \mu_i + e_i,$$

$$e_i \sim N(0,\ \sigma_i^2), (1)$$

where $e_i$ represents within-study sampling error corresponding to the *i*-th estimate $y_i$, and $\sigma_i^2$ is the sampling variance, typically estimated from study data and treated as known. This assumption of fixed $\sigma_i^2$, despite some uncertainty in small samples, distinguishes meta-analysis from standard (weighted) linear regression. According to the assumption of the RE model, the true effect size parameter $\mu_i$ are assumed to vary across studies due to heterogeneity (Borenstein et al., 2009):

$$\mu_i = \mu + u_i,$$

$$u_i \sim N(0, \tau^2), (2)$$

Where $\mu$ denotes the average true effect $E(\mu_i)$ (also known as grand mean or overall mean), $u_i$ captures the study-specific deviation from $\mu$, and $\tau^2$ quantifies the between-study variance, often termed heterogeneity. This hierarchical structure allows the RE model to accommodate variability in true effects, a critical feature in preclinical meta-analysis where study conditions differ widely (Ineichen et al., 2024).

When $\tau^2 = 0$, the RE model reduces to the fixed-effect (or common-effect) model, assuming (i.e., $\mu \equiv \mu_i$). When $\tau^2 \neq 0$, these exists random or systematic differences in study characteristics (e.g., population, treatment, and outcomes). To account for systematic heterogeneity, the mixed-effects meta-regression model extends the RE framework by incorporating predictor variables ("moderators"):

$$\mu_i = \beta_0 + \beta_1 x_{i1} + \cdots + \beta_p x_{ip} + u_i,$$

$$u_i \sim N(0, \tau^2), (3)$$

Where $x_{i1}, \ldots, x_{ip}$ are the values of moderator variables ((e.g., species, dosage, or study quality)), $\beta_{i1}, \ldots, \beta_{ip}$ are regression coefficients estimating the impact of these moderators on $\mu_i$, and $\tau^2$ now represents he amount of "residual heterogeneity", capturing the variability in the true effect parameters $\theta_i$ not accounted for by the moderator variables $x_{i1}, \ldots, x_{ip}$. Both quantitative and qualitative moderators (via dummy coding) are possible, and modern meta-analytic software (R packages metafor and brms) supports complex specifications (Bürkner, 2017; Viechtbauer, 2010), such as interaction terms or nonlinear relationships modelled through polynomial or spline functions.

The RE and mixed-effects models impose several assumptions, some explicit and others implicit (Jackson & White, 2018), which constrain the distributional parameters and regression structure. Violations of these assumptions may compromise inference. Below, we outline these assumptions (Table 1), leveraging consistent notation to facilitate comparison with the distributional regression framework introduced later.

Table 1

Summary of assumption violations in statistical models for the conventional meta-analysis and existing modelling approaches.

| Assumptions | Violation | Existing modelling approaches |
| --- | --- | --- |
| Normal sampling errors | Non-normal data (e.g., binary or count outcomes) | Generalized linear mixed effects models |
| Dependent effect sizes | Nested or correlated effect sizes | Multilevel and multivariate models |
| Homogeneous variance | Heterogeneity varies by study characteristics | Location-scale models |
| Normal random effects | Outliers or non-replicable results | Robust models |

## 2.1 Normality of within-study sampling errors

One implicit assumption in the RE model is that within-study sampling errors ($e_i$) are normally distributed (Jackson & White, 2018). For continuous outcomes, such as standardized mean differences, this assumption holds by invoking the central limit theorem, which ensures approximate normality for sufficiently large samples. However, in preclinical research, where binary outcomes (e.g., presence or absence of a disease phenotype) can exhibit skewed or heavy-tailed distributions, including risk differences ($RD = p_1 - p_2$), odds ratios ($OR = \frac{p_1/(1-p_1)}{p_2/(1-p_2)}$),), or risk ratios ($p_1/p_2$, where $p_1$ and $p_2$ denotes event probabilities). Similarly, count data (e.g., lesion counts in pathology studies) follow binomial or Poisson distributions, not normal ones. Such misspecification can undermine the nominal properties of confidence intervals and hypothesis tests, potentially leading to inflated Type I errors or biased estimates (Bhaumik et al., 2012; Stijnen et al., 2010). To address this violation, generalized linear mixed models (GLMMs) have been proposed, which replace the normal distribution with one tailored to the outcome type (e.g., binomial for binary data, Poisson for counts) (Jackson et al., 2018; Jansen & Holling, 2023; Stijnen et al., 2010).

**2.2 Independence of effect sizes**

Another common assumption in the RE models is that each study contributes a single, independent effect size. However, recent surveys highlight that preclinical meta-analyses frequently encounter effect size multiplicity (López-López et al., 2018), where studies report multiple related effect sizes, introducing dependencies among both observed values and their true counterparts (Yang et al., 2022). Such dependencies arise when studies measure multiple outcomes (e.g., weight and cholesterol levels in metabolic studies), assess effects at different time points (e.g., behavioural responses at 24 and 48 hours), or use shared experimental units, leading to correlated errors or true effects. Ignoring these correlations underestimates standard errors, potentially inflating Type I error rates and compromising the validity of pooled estimates (Van den Noortgate et al., 2013). To accommodate hierarchical dependencies, multilevel meta-analysis (MLMA) models introduce nested random effects to preclinical meta-analysis (Yang et al., 2022). For dependencies among correlated outcomes, multivariate meta-analysis (MVMA) models the joint distribution of multiple effect sizes (Riley, 2009). In a meta-analysis of neuroprotective drugs, where studies report SMDs for infarct volume across multiple time points, an MLMA model captures within-study correlations, while an MVMA model jointly estimates correlated endpoints like lesion size and motor function, improving efficiency and accuracy.

### 2.3 Homogeneous between-study heterogeneity

Recall Equations 2 and 3, the (residual) between-study heterogeneity $\tau^2$ is assumed to be a fixed variance parameter. This assumption of homogeneous heterogeneity implies that variability in true effects is uniform, regardless of study characteristics. However, in preclinical research, heterogeneity may vary systematically due to factors such as species, experimental design, or study quality. For example, a meta-analysis of 539 randomized controlled trials on psychotherapy for depression identified geographical region as a significant predictor of heterogeneity magnitude, suggesting that $\tau^2$ is not fixed but context-dependent (Kuper et al., 2025). To address this violation, location-scale models have been recently introduced to extend the RE framework by allowing the heterogeneity parameter to vary systematically as a function of moderator variables. Location-scale models have been recently introduced to the context of meta-analysis (Blázquez-Rincón et al., 2025; Röver & Friede, 2023; Viechtbauer & López-López, 2022).

### 2.4 Normality of between-study random effects

Beyond assuming normality for within-study errors, the RE model posits that true effect sizes follow a normal distribution: $\mu_i \sim N(\mu, \tau^2)$ (Equation 2), implying a symmetric, unimodal distribution with no extreme outliers or structural zeros beyond what random variation predicts. Extreme effect sizes can arise due to unique study designs or small sample sizes in preclinical research. For example, a rodent study with an unusually large effect size (e.g., due to an atypical strain) may lie in the heavy tails of the distribution, skewing the pooled estimate and inflating $\tau^2$. This suggests a non-normal, heavy-tailed distribution for $\mu_i$. Similarly, in studies of rare events (e.g., adverse reactions in toxicology studies), many effect sizes may be zero, reflecting either true null effects or insufficient power to detect effects. This creates a point mass at zero, incompatible with a continuous normal distribution, and can bias the pooled estimate toward zero. Recently, robust meta-analysis models address outliers by using heavy-tailed distributions (e.g., the $t$-distribution) for $\mu_i$, reducing the influence of extreme values and stabilizing estimates (Wang et al., 2025). For excess zeros, zero-inflated meta-analysis models employ a mixture model that separates structural zeros (e.g., true absence of effect) from non-zero effects, providing a more accurate representation of the data (Li et al., 2025).

## 3 The distributional regression framework for meta-analysis

Distributional regression models, which allow all parameters of a response distribution to depend on predictors, provide a flexible framework for statistical modelling. Notable implementations, such as generalized additive models for location, scale, and shape (M. D. Stasinopoulos et al., 2024), are well-documented in statistical software such as R (Rigby et al., 2019; D. M. Stasinopoulos & Rigby, 2008). While these models are widely used in various fields, their potential in meta-analysis remains underexplored. In meta-analysis, traditional models focus solely on the mean effect size, often overlooking heterogeneity in other distributional properties like scale or shape. The proposed distributional regression framework addresses this limitation by modelling the full distribution of effect sizes, allowing all parameters—location, scale, and shape—to vary across studies as functions of moderators and random effects. This approach unifies and extends conventional meta-analytic models, including random-effects, multilevel, multivariate, location-scale, and robust models, under a single framework.

Consider $y_{ij}$ as the $j$-th observed effect size from the $i$-th study, where $i = 1, \ldots, k$ indexes studies and $j = 1, \ldots, n_i$ indexes effect sizes within each study. We assume that $y_{ij}$ follows a probability distribution with density (or mass) function:

$$y_{ij} \sim f(y_{ij} | \theta_{ij}), (4)$$

Where $f$ is the probability density function for continuous distributions (e.g., normal) or probability mass function for discrete distributions (e.g., Poisson, negative binomial), $\theta_{ij} = (\theta_{ij}^{(1)}, \theta_{ij}^{(2)}, \ldots, \theta_{ij}^{(p)})^\text{T}$ is a vector of $p$ parameters defining the distribution's properties (e.g., mean, variance, skewness). For each parameter $m = 1, \ldots, p$, we model $\theta_{ij}^{(m)}$ using a mixed-effects regression structure:

$$g_m\left(\theta_{ij}^{(m)}\right) = \mathbf{x}_{ij}^\text{T} \boldsymbol{\beta}^{(m)} + \mathbf{z}_{ij}^\text{T} \mathbf{u}_i^{(m)}, (5)$$

where $g_m$ is the link function specific to the $m$-th parameter (e.g., identity for location, log for scale), $\mathbf{x}_{ij}$ is a vector of moderator variables associated with fixed effects, $\boldsymbol{\beta}^{(m)}$ is a vector of fixed-effect regression coefficients for the $m$-th parameter, $\mathbf{z}_{ij}$ is vector of covariates for random effects, $\mathbf{u}_i^{(m)}$ is a vector of random effects for the $i$-th study capturing unexplained heterogeneity in the $m$-th parameter, typically $\mathbf{u}_i^{(m)} \sim N(\mathbf{0}, \mathbf{\Sigma}^m)$. Different meta-analysis models can be represented as special versions of the general framework in Equations 4 to 5. We depict the most common ones in Figure 1 and highlight their connection to the general framework.

**4 Special cases: Unifying existing meta-analytic models**

The distributional regression framework offers a comprehensive and flexible approach to meta-analysis, subsuming a diverse range of existing models under a unified structure. This section demonstrates how it encompasses (1) the standard RE model and (2) advanced models relaxing assumptions summarized in Table 1. These are illustrated with preclinical research examples, where such challenges are common due to experimental designs and data complexity. Through targeted constraints on the probability density function $f$, the number of parameters $p$, and the regression specification, each model emerges as a special case, tailored to specific data challenges prevalent in preclinical research and beyond.

**4.1 The standard random-effects model**

The standard RE model, a cornerstone of meta-analysis, requires that each study $i$ provides a single effect size $y_i$ (e.g., a standardized mean difference or log-odds ratio) with known within-study variance (Borenstein et al., 2009). This model emerges from the distributional regression framework through the following constraints. Within the distributional regression framework, this model is specified by setting the effect size distribution to:

$$y_i \sim f = N(\mu_i, \sigma_i^2), (6)$$

with $p = 1$ distributional parameter, $\theta_i^{(1)} = \mu_i$ as the location parameter. An identity link is applied as $g_1(\theta_i^{(1)}) = \theta_i^{(1)} = \mu_i$. The location parameter is modelled as $\mu_i = x_i^T \beta^{(1)} + z_i^T u_i^{(1)}$, where $x_i = 1$, $\beta^{(1)} = \mu$ (the overall mean), $z_i = 1$, and $u_i^{(1)} = u_i \sim (0, \tau^2)$. Taken together, the above specification yields the familiar marginal form $y_i \sim (x_i^T \beta^{(1)}, \tau^2 + \sigma_i^2)$.

**4.2 Generalized linear mixed models for non-normal data**

For effect sizes with non-normal distributions, e.g., binary or count outcomes in preclinical studies, GLMMs provide a robust solution (Jackson et al., 2018; Y. Lee & Nelder, 2006; Stijnen et al., 2010). The explicit constraint to a single parameter with a binomial distribution and logit link ensures the framework's adaptability to discrete data. Consider a meta-analysis of therapeutic response rates in animal models, where responses follow a binomial distribution:

$$y_{ij} \sim f = \text{binomial}(n_{ij}, p_{ij}), (7)$$

with success probability $p_{ij}$. Using a logit link leads to $g_1(\theta_{ij}^{(1)}) = \text{logit}(p_{ij})$. the regression structure becomes $\text{logit}(p_{ij}) = x_{ij}^T \beta^{(1)} + z_{ij}^T u_i^{(1)}$, where $x_{ij} = 1$, $\beta^{(1)} = \beta$ (fixed effect), $z_{ij} = 1$, $u_i^{(1)} = \mu_i \sim (0, \tau^2)$ (random effects). This specification results in $\text{logit}(p_{ij}) = \beta + u_i$. Moderator variables can be added to the model to explain heterogeneity.

Another common case involves count data, such as lesion counts in animal pathology studies. Here, the effect sizes follow a Poisson distribution:

$$y_{ij} \sim f = \text{Poisson}(\lambda_{ij}), (8)$$

with $\lambda_{ij}$ being the expected count for observation $j$ in study $i$. Applying the log link, $g_1(\theta_{ij}^{(1)}) = \log(\lambda_{ij})$, to ensure the expected count $\lambda_{ij}$ remains positive. The regression structure is $\log(\lambda_{ij}) = x_{ij}^T \beta^{(1)} + z_{ij}^T u_i^{(1)}$, where $x_{ij} = 1$, $\beta^{(1)} = \beta$ (fixed effect), $z_{ij} = 1$, $u_i^{(1)} = u_i \sim (0, \tau^2)$ (random effects). The above specification can be simplified to $\log(\lambda_{ij}) = \beta + u_i$.

**4.3 Multilevel models for hierarchical data**

In meta-analyses where studies report multiple effect sizes organized hierarchically, MLMA models offer a sophisticated approach to account for dependencies across levels (Riley, 2009; Van den Noortgate et al., 2013). Such structures are common in fields like preclinical research (Yang et al., 2022). Let $y_{ij}$ represent the observed effect size for the $j$-th effect in the $i$-th study nested or cross-classified within the $s$-th cluster (e.g., research labs or experimental paradigms). The MLMA model is defined as:

$$y_{sij} \sim f = N(\mu_{sij}, \sigma_{sij}^2), (8)$$

with $\mu_{sij} = x_{sij}^T \beta + u_s + v_{si} + \omega_{sij}$, where the $x_{sij}^T \beta$ denotes the fixed effects with $x_{sij}$ being moderator variables and $\beta$ being the corresponding regression coefficients, $u_s$ denotes the cluster-level random effects following a normal distribution $N(0, \tau_u^2)$ (where $\tau_u^2$ captures cluster-level variation), $v_{si}$ denotes the study-level random effects following a normal distribution $N(0, \tau_v^2)$ (where $\tau_u^2$ captures study-level variation), and $w_{sij}$ denotes effect-size level random effects following a normal distribution $N(0, \tau_\omega^2)$ (where $\tau_\omega^2$ captures effect-level variation).

The MLMA model fits seamlessly into the distributional regression framework by specifying setting $f = N(\mu_{sij}, \sigma_{ij}^2)$, with $p = 1$, $\theta_{sij}^{(1)} = \mu_{sij}$, and an identity link $g_1(\theta_{sij}^{(1)}) = \mu_{sij}$. The mean parameter is expressed as $\mu_{sij} = x_{sij}^T \beta^{(1)} + z_{sij}^T u_{sij}^{(1)}$, where $z_{sij}^T = [1,1,1]$ aligns with the three levels of random effects, and $u_{sij}^{(1)} = [u_s, v_{si}, \omega_{sij}]^T \sim N(\mathbf{0}, \text{diag}(\tau_u^2, \tau_v^2, \tau_\omega^2))$ captures the hierarchical random effects.

**4.4 Multivariate models for correlated data**

The MVMA model addresses scenarios where studies report multiple correlated outcomes (Cheung, 2013; Jackson et al., 2011), such as behavioural and physiological measures in preclinical research. The distributional regression framework provides a unified structure for modelling these outcomes by specifying the joint distribution of the outcome vector $\mathbf{y}_i = (y_{i1}, \ldots, y_{il})^T$, where $l$ denotes the number of outcomes per study $i$. For clarity, we assume a multivariate normal distribution:

$$\mathbf{y}_i \sim f = N(\boldsymbol{\mu}_i, \boldsymbol{\Sigma}_i), (9)$$

where $\boldsymbol{\mu}_i = (\mu_{i1}, \ldots, \mu_{il})^T$ is the mean parameter vector, and $\boldsymbol{\Sigma}_i$ is the within-study sampling variance-covariance matrix. While $\boldsymbol{\Sigma}_i$ is typically assumed to be known from study data, in practice we have to impute it (e.g., assuming an arbitrary constant sampling correlation coefficient like 0.5 or 0.7) or estimate it based on study data. Each mean parameter is modelled as $\mu_{ik} = x_{ik}^T \beta^{(k)} + z_{ik}^T u_i^{(k)}$, for $k = 1, \ldots, l$, with $p = l$, and random effects $\mathbf{u}_i = (u_i^1, \ldots, u_i^l)^T \sim N(\mathbf{0}, \mathbf{T})$, where $\mathbf{T}$ is the between-study variance-covariance matrix. To enhance the illustration, consider a bivariate case, where each study reports two correlated outcomes. The outcome vector is $\mathbf{y}_i =$ $(y_{i1}, y_{i2})^T$, with mean parameters $\boldsymbol{\mu}_i = (\mu_{i1}, \mu_{i2})^T$. The between-study variance-covariance matrix is specified as $\mathbf{T} = \begin{pmatrix} \tau_1^2 & \rho\tau_1\tau_2 \\ \rho\tau_1\tau_2 & \tau_2^2 \end{pmatrix})$, where $\rho$ is the between-study correlation.

**4.5 Location-scale models for heteroscedasticity**

In meta-analysis, the assumption of homogeneous between-study variance could fail (Table 1). The distributional regression framework addresses this through location-scale models, which simultaneously model the mean and variance of effect sizes, allowing the between-study variance to vary with study-level covariates. This approach extends the traditional random-effects model by incorporating a heterogeneous variance structure (Viechtbauer & López-López, 2022). For the location-scale model, the distribution is specified as:

$$y_i \sim f = N(\mu_i, \sigma_i^2), (10)$$

where the mean parameter is $\mu_i = x_i^T \beta^{(1)} + z_i^T u_i^{(1)}$ with $u_i^{(1)} \sim N(0, \tau_i^2)$. The study-specific between-study variance is modelled as

$$\log(\tau_i^2) = x_i^T \beta^{(2)} + z_i^T u_i^{(2)}, (11)$$

where $u_i^{(2)} \sim N(0, \tau_\tau^2)$ captures residual heterogeneity in the Equation 11. Here, the number of distributional parameters $p$ is set to 2, with $\theta_i^{(1)} = \mu_i$ and $\theta_i^{(2)} = \tau_i^2$, using an identity link for the mean and a log link for the variance to ensure positivity. This structure captures heteroscedasticity and can be extended to multilevel contexts by incorporating hierarchical random effects (see "4.3 Multilevel models for hierarchical data").

**4.6 Robust models for outliers**

Outliers and extreme effect sizes pose a significant challenge in meta-analysis, particularly in preclinical studies with small sample sizes, where extreme effect sizes can skew pooled estimates. The true effects can exhibit heavy tails or skewness, as seen in studies with atypical designs or populations (Baker & Jackson, 2016). Robust models within the distributional regression framework employ long-tailed, skewed, or mixture distributions to reduce the influence of outliers. For instance, the random effects can be modelled using a $t$-distribution to accommodate heavier tails (K. J. Lee & Thompson, 2008):

$$\mu_i \sim t(\mu, s, v), (12)$$

where $s = \sqrt{\tau^2 \frac{(v-2)}{v}}$, and $v$ is the degrees of freedom determining the weight of the tails. Alternatively, a skewed $t$-distribution can capture asymmetry (Panagiotopoulou et al., 2024):

$$\mu_i \sim t(\xi, s, v, \gamma), (13)$$

where $\xi$, and $s$ are location, and scale parameters, $v$ and $\gamma$ are parameters to control the shape of the distribution. The beta distribution of the random effects is restricted to a bounded interval, which leads to a short-tailed distribution for meta-analyses with no outliers at all. The between-study distribution can be modelled as

$$\mu_i \sim \text{Beta}(a, b), (14)$$

where $a = \mu(\frac{\mu(1-\mu)-\tau^2}{\tau^2})$ and $b = (1-\mu)(\frac{\mu(1-\mu)-\tau^2}{\tau^2})$, assuming $a, b > 1$. To model data from multiple subpopulations or account for outlying studies, a mixture of normal distributions can be used (Panagiotopoulou et al., 2024):

$$\mu_i \sim w_1 N(\mu_1, \tau_1^2) + \cdots + w_1 N(\mu_n, \tau_n^2), (15)$$

where $\mu_1, \ldots, \mu_n$ and $\tau_1^2, \ldots, \tau_n^2$ are the subgroup-specific mean and variance for the $n$ subgroups, $w_1, \ldots, w_n$ are corresponding mixing weights.

**5. Case Study: Small-Study effects and heterogeneity in 67,393 meta-analyses**

There has long been debate surrounding the limitations of small trials in evidence synthesis. One widely studied issue is the "small-study effect" where smaller studies tend to report larger treatment effects, which is often interpreted as evidence of publication bias (Egger et al., 1997). A less frequently tested hypothesis concerns whether smaller studies also exhibit greater between-study heterogeneity (IntHout et al., 2015; Stanley et al., 2022). If confirmed, this would challenge the foundational assumption of the conventional random-effects meta-analysis model, which presumes that heterogeneity (typically denoted as tau-squared) is independent of sampling error (see Equation 2).

To illustrate the utility of distributional regression for meta-analysis, we reanalysed 67,393 meta-analyses from the Cochrane Database of Systematic Reviews, compiled by Bartoš et al. (2024), using a location-scale model. In this framework, both the average treatment effect (location) and the between-study heterogeneity (scale) are modelled as functions of a covariate: the standard error of effect size estimates (Nakagawa et al., 2025; Viechtbauer & López-López, 2022). Since standard error is inversely related to within-study sample size, it serves as a proxy for study precision or trial size. First, we applied an original version of Egger's regression test, regressing the treatment effect on the standard error. Out of 67,393 meta-analyses, 8,766 (13.0%) showed evidence that smaller studies reported significantly larger effect sizes. Second, we fitted a full location-scale model that simultaneously included standard error as a predictor of both the treatment effect and the heterogeneity parameter. This model identified 6,897 meta-analyses (10.2%) that showed evidence that reported significantly larger effect sizes, and 1,519 meta-analyses (2.3%) showed evidence that smaller studies were associated with greater heterogeneity.

These findings highlight the nuanced patterns of small-study effects that traditional meta-analytic approaches may miss. Conventional models typically examine either the average effect or the presence of publication bias in isolation, and often assume homogeneity or a constant heterogeneity. In contrast, the distributional regression framework enables simultaneous modelling of multiple distributional parameters, providing richer insights into the structure of bias and heterogeneity in meta-analytic datasets. However, these results should be interpreted with caution and are intended for illustrative purposes only. Regression-based methods for detecting publication bias, including Egger's test and its extensions, are known to have limited statistical power (Carter et al., 2019), particularly when applied to meta-analyses with a small number of studies. This limitation is especially relevant here, as the Cochrane Database of Systematic Reviews typically includes a median of only five studies per meta-analysis (Bartoš et al., 2024). Moreover, publication bias and heterogeneity may be influenced by additional factors not captured in the current modelling framework.

## 6. Discussion

The distributional regression framework for meta-analysis introduced in this paper offers a broad and integrated perspective on statistical modelling in evidence synthesis. It extends standard meta-analytic models by allowing every parameter of the effect size distribution, such as the average (location), variability (scale), and distributional shape, to vary across studies and to be modelled as a function of study-level characteristics. This framework can represent many existing models, including random effects models, multilevel models, multivariate models, and location-scale models, as special cases. In doing so, it provides conceptual clarity for the connection between different existing models for meta-analysis. Importantly, it also encourages the formulation of new types of questions that have been largely neglected in meta-analytic research, such as why some studies yield more variable effects than others or whether the shape of the effect size distribution changes across contexts.

**Software implementation**

Currently, no dedicated software package has been developed specifically for distributional regression in meta-analysis. Nonetheless, several existing tools can be adapted to support this modelling framework. In the Bayesian context, the brms package in R provides robust support for modelling multiple distributional parameters through Stan's probabilistic programming language as backend (Bürkner, 2017). It can incorporate known sampling variances via analytic weights and supports a wide variety of outcome distributions, making it a suitable platform for fitting distributional regression for meta-analysis.

In addition, special cases of distributional regression for meta-analysis discussed in this paper can be implemented using current meta-analytic packages. For example, metafor can handle generalized linear models for meta-analysis (Viechtbauer, 2010), metaSEM supports three-level and structural equation meta-analysis (Cheung, 2015), and mixmeta provides tools for multivariate models (Sera et al., 2019). The blsmeta package implements location-scale models (Rodriguez et al., 2023), while metaplus allows for non-normal or mixture distributions for random effects (Beath, 2016). Although these packages often assume normality or are limited to a subset of parameters, they can serve as building blocks or be extended to fit within the broader framework of distributional regression.

Beyond meta-analysis, several packages implement general distributional regression approaches that could be adapted to accommodate known sampling error and study weights. The gamlss package supports flexible modelling of multiple distributional parameters using penalized likelihood inference (D. M. Stasinopoulos & Rigby, 2008). Similarly, bamlss provides a Bayesian approach based on Markov chain Monte Carlo (Umlauf et al., 2021). These platforms offer valuable infrastructure for the future development of software explicitly designed for meta-analysis within the distributional regression framework.

**Model selection and pre-registration**

The flexibility of distributional regression models necessitates rigorous model specification practices. Because the framework allows each parameter of the distribution to be linked to covariates, it is essential that researchers specify in advance which parameters are of interest and which covariates are expected to explain their variation. Hypotheses and model structures should ideally be defined based on prior knowledge, rather than data-driven exploration. This is especially important when moving beyond the mean, since the empirical basis for studying variance or shape parameters remains underdeveloped in many fields. We therefore encourage researchers to pre-register not only their primary hypotheses but also the planned model structures for all relevant distributional parameters.

In cases where multiple plausible models exist, formal model selection criteria can assist in comparing alternative specifications. In the frequentist setting, model selection supports the use of information criteria such as AIC and BIC(Brooks et al., 2017), while in the Bayesian setting, brms provides tools for approximate leave-one-out cross-validation and other model evaluation metrics (Vehtari et al., 2017). Visual diagnostics and posterior predictive checks can further aid in identifying model misspecifications. Despite these tools, formal guidance is still limited for model selection in distributional regression meta-analysis, particularly when accounting for sampling error and dependence structures. We call on statistical methodologists to investigate and improve these tools in the context of evidence synthesis.

**Limitations**

Several limitations of the proposed framework should be acknowledged. First, while distributional regression for meta-analysis allows the full distribution of effect sizes to be modelled using covariates, most applied research questions and hypotheses in the meta-analytic literature continue to focus on average effects. This emphasis on mean parameters may limit the practical adoption of distributional regression in its full generality. For the framework to realize its potential, researchers must begin to formulate and test hypotheses about the variability and shape of effect size distributions, not just their central tendencies.

Second, the flexibility of the framework comes at the cost of increased model complexity. Allowing multiple distributional parameters to vary with covariates can rapidly increase the number of model parameters, especially in cases where multivariate or hierarchical structures are included. For example, multivariate meta-analytic models are known to be highly parameterized, and extending these models to allow multiple distributional parameters to vary may lead to overparameterization. This, in turn, can compromise statistical power, model convergence, and interpretability (Boca et al., 2017). As with any complex modelling framework, the benefits of flexibility must be weighed against the risks of model overfitting and reduced robustness. Careful model selection, validation, and theoretical justification are essential to mitigate these risks.

## 7. Conclusion and future perspective

Meta-analysis has traditionally centred on estimating average effect sizes while assuming fixed distributional forms. This paper introduces a more expansive framework, distributional regression for meta-analysis, that enables researchers to model not only the mean but also the variance and shape of effect size distributions as functions of covariates. This generalization unifies a wide range of existing models and allows for the investigation of previously inaccessible questions about the sources and nature of heterogeneity in scientific research. We hope the proposed framework will inspire researchers to move beyond mean-centric thinking and to explore how methodological and biological factors shape the full distribution of effect sizes. In our own case study, we examined whether smaller studies report not only larger effects but also less variability, illustrating how hypotheses about multiple distributional parameters can be tested simultaneously.

Importantly, the proposed approach encourages the formulation and testing of hypotheses related to distributional parameters beyond the mean, such as scale and shape. In several other fields, such hypotheses have already been empirically tested, demonstrating the scientific value of modelling distributional properties. For example, research in ecology and evolution has shown that phenotypic skewness in body size among juvenile birds (blue tits) is driven by environmental factors, highlighting how distributional asymmetries carry biological significance (Pick et al., 2022). The relationship between the skewness and the kurtosis of functional trait distributions has been quantified to detect ecological thresholds (Le Bagousse-Pinguet et al., 2025). Similarly, studies in quantitative genetics have explored how genetic, mutational, and environmental variances vary and covary across traits, offering deeper insights into trait evolution (Hansen et al., 2011; King et al., 2025). These examples illustrate the utility of distributional modelling and suggest that preclinical science can benefit by borrowing these successful approaches to generate and test richer hypotheses using the proposed framework. As evidence synthesis becomes increasingly important in guiding policy, practice, and theory, tools that provide a deeper understanding of heterogeneity and bias are urgently needed. We encourage the meta-analytic community to explore the possibilities afforded by distributional regression, to formulate richer hypotheses, and to contribute to the ongoing development of methods and software that support this next generation of meta-analytic modelling.

**Acknowledgments**

YY, and SN were supported by the Australian Research Council Discovery Grant (DP210100812 & DP230101248). SN acknowledges support from a Canada Excellence Research Chair (CERC-2022-00074).

**Author contributions**

Yefeng Yang: Conceptualization; formal analysis; investigation; methodology; software; writing – original draft; writing – review and editing. SN: Conceptualization; investigation; methodology; writing – review and editing; funding acquisition; supervision. All authors approved the final manuscript.

**Competing interests**

All authors declare no competing interests.